\documentclass[aps,prl,reprint,twocolumn,superscriptaddress,floatfix,nofootinbib,longbibliography]{revtex4-2}
\usepackage{graphicx,amsmath,amsfonts,amssymb,amsthm,xr}
\usepackage{epsfig,amsmath,amssymb,color,dsfont,upgreek,physics}
\usepackage{mathrsfs}
\usepackage{mathtools}
\usepackage{bbold}
\usepackage{float}
\usepackage[caption=false]{subfig}
\usepackage[bookmarks=true,colorlinks,linkcolor=OrangeRed,urlcolor=NavyBlue,citecolor=RoyalBlue]{hyperref}
\usepackage[capitalize]{cleveref}
\usepackage[dvipsnames]{xcolor}
\usepackage{orcidlink}

\newcommand{\hpa}{h_{\parallel}}
\newcommand{\hpe}{h_{\perp}}

\usepackage{graphicx}
\usepackage{dcolumn}
\usepackage{bm}
\usepackage{color}

\newcommand{\rr}{\mathbf{r}}

\begin{document}

\title{Microscopic Dynamics of False Vacuum Decay in the $2+1$D Quantum Ising Model}

\author{Umberto Borla${}^{\orcidlink{0000-0002-4224-5335}}$}
\affiliation{Max Planck Institute of Quantum Optics, 85748 Garching, Germany}
\affiliation{Department of Physics and Arnold Sommerfeld Center for Theoretical Physics (ASC), Ludwig Maximilian University of Munich, 80333 Munich, Germany}
\affiliation{Munich Center for Quantum Science and Technology (MCQST), 80799 Munich, Germany}

\author{Achilleas Lazarides${}^{\orcidlink{0000-0002-0698-2776}}$}
\affiliation{Loughborough University, Loughborough, Leicestershire LE11 3TU, UK}

\author{Christian Gro\ss${}^{\orcidlink{0000-0003-2292-5234}}$}
\affiliation{Physikalisches Institut and Center for Integrated Quantum Science and Technology, Eberhard Karls Universit\"at T\"ubingen, 72076 T\"ubingen, Germany}

\author{Jad C.~Halimeh${}^{\orcidlink{0000-0002-0659-7990}}$}
\email{jad.halimeh@lmu.de}
\affiliation{Department of Physics and Arnold Sommerfeld Center for Theoretical Physics (ASC), Ludwig Maximilian University of Munich, 80333 Munich, Germany}
\affiliation{Max Planck Institute of Quantum Optics, 85748 Garching, Germany}
\affiliation{Munich Center for Quantum Science and Technology (MCQST), 80799 Munich, Germany}
\affiliation{Department of Physics, College of Science, Kyung Hee University, Seoul 02447, Republic of Korea}

\date{\today}

\begin{abstract}
False vacuum decay, which is understood to happen through bubble nucleation, is a prominent phenomenon relevant to elementary particle physics and early-universe cosmology. Understanding its microscopic dynamics in higher spatial dimensions is currently a major challenge and research thrust. Recent advances in numerical techniques allow for the extraction of related signatures in tractable systems in two spatial dimensions over intermediate timescales. Here, we focus on the $2+1$D quantum Ising model, where a longitudinal field is used to energetically separate the two $\mathbb{Z}_2$ symmetry-broken ferromagnetic ground states, turning them into a ``true'' and ``false'' vacuum. Using tree tensor networks, we simulate the microscopic dynamics of a spin-down domain in a spin-up background after a homogeneous quench, with parameters chosen so that the domain corresponds to a bubble of the true vacuum in a false-vacuum background. Our study identifies how the ultimate fate of the bubble---indefinite expansion or collapse---depends on its geometrical features and on the microscopic parameters of the Ising Hamiltonian. We further provide a realistic quantum-simulation scheme, aimed at probing bubble dynamics on atomic Rydberg arrays. 
\end{abstract}

\maketitle

\textbf{\textit{Introduction.---}}
The problem of vacuum metastability plays a prominent role in modern particle physics and early-universe cosmology \cite{turner_is_1982, coleman_fate_1977, callan_fate_1977, coleman_gravitational_1980, Coleman_1985, vilenkin1994}. The current Standard Model predictions, indeed, hint at the fact that the electroweak vacuum might be a local minimum, separated by the true vacuum of our universe by a finite energy barrier, and may therefore decay over extremely large timescales \cite{isidori_metastability_2001, degrassi_higgs_2012}. Independently of the particular model under consideration, the decay of a false vacuum is understood to happen through a process of bubble nucleation \cite{LANGER1967108, Kobzarev:1974cp, coleman_fate_1977}, where bubbles of the true vacuum appear as fluctuations of a metastable ground state. If one such bubble is larger than a certain critical size it can expand indefinitely, causing the true vacuum to take over.

\begin{figure}[t!]
    \centering
    \includegraphics[width=\linewidth]{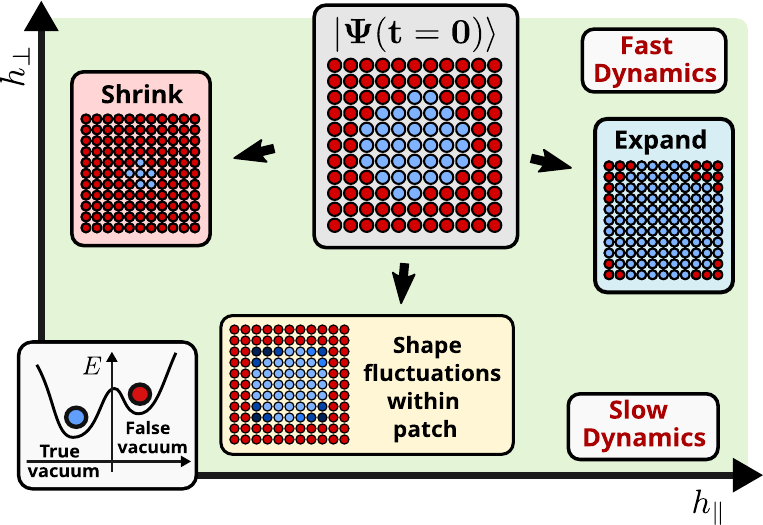}
    \caption{Sketch of the dynamical phase diagram corresponding to a homogeneous quench, in the ordered phase, for an initial state $|\Psi(t=0)\rangle$ consisting of a domain of spin-downs (true vacuum) in a spin-up background (false vacuum). The quench is performed with the Ising Hamiltonian \eqref{eq:H_TLFI} with parameters $\hpe$ and $\hpa$. The transverse field $\hpe$ determines the speed of the dynamics. Even at small $\hpe$ shapes with the same perimeter can freely fluctuate on short timescales within the square in which they are inscribed, as there is no surface-energy cost to this. This is purely a lattice effect. The longitudinal field $\hpa$, on the other hand, sets the energy difference between the two vacua and favors bubble expansion.}
    \label{fig:fig1}
\end{figure}

Direct simulation of real-time false-vacuum phenomenology is notoriously challenging due to its non-perturbative character and the very large timescales involved \cite{voloshin1985id, Rutkevich1999, lagnese_false_2021, johansen2025manybodytheoryfalsevacuum}, even in the most simplified scenarios. However, recent advances in tensor network simulations \cite{montangero2018introduction,Schollwoeck2011_review,Paeckel2019_review,Orus2019TensorNetworks} have rendered the task more accessible, and several works have shed light on bubble nucleation and signatures of vacuum metastability after quantum quenches in $1+1$D (one spatial and one temporal dimensions)\footnote{We denote a system or phenomenon in $d$ spatial dimensions as $d+1$D or $d$d.} \cite{lagnese_false_2021, lagnese_detecting_2024, maertens2025realtimebubblenucleationgrowth}. The simulation of long-time dynamics in $d>1$ spatial dimensions still faces considerable hurdles, but tree tensor network (TTN) \cite{tagliacozzo_simulation_2009, murg_simulating_2010} representations of the wave function of the system have recently proven very effective in accessing intermediate timescales. This has been exploited to study interface dynamics, melting, and scattering in the $2+1$D quantum Ising model \cite{balducci_localization_2022, hart_hilbert_2022, balducci_interface_2023,Krinitsin2025TimeEvolution, pavesic_constrained_2025, pavesic_scattering_2025}. 

At the same time, there is a significant effort towards investigating such phenomena on a variety of quantum simulation platforms \cite{Labuhn2016Tunable,darbha_false_2024,Zenesini2024FalseVacuumDecay, zhu_probing_2024,Kavaki2025FalseVaccumDecay, manovitz_quantum_2025, vodeb_stirring_2025, chao_probing_2025, osterholz_collective_2025}. Due to its universal interest and versatility, the implementation of the quantum Ising model is of utmost importance and has been pursued in several settings, most notably on programmable Rydberg atom arrays. Future experiments are expected to realize sufficiently large $2$d lattices and accurately simulate quantum dynamics over long timescales, paving the way for the observation of vacuum-decay physics on quantum hardware.

In this Letter, we provide a detailed study of the dynamics of true vacuum bubbles in a false-vacuum background after a global quench. We show how an interplay of quench parameters, linear size, and shape of the bubble determines whether it expands or shrinks, clarifying what makes a bubble critical in a non-perturbative regime characterized by strong quantum fluctuations. To conclude, we outline a feasible experimental quantum-simulation proposal, whose realization will be guided by the concrete results that we provide here. 

\textbf{\textit{Model.---}}
We consider the quantum Ising model with transverse and longitudinal fields (TLFIM) on a square lattice with unity spacing, described by the Hamiltonian
\begin{equation}
    \hat{H} = -J \sum_{\rr, \eta=\hat{x},\hat{y}} \hat\sigma^x_\rr \hat\sigma^x_{\rr+\eta} - \hpe \sum_\rr \hat\sigma^z_\rr - \hpa \sum_{\rr} \hat\sigma^x_\rr,
    \label{eq:H_TLFI}
\end{equation}
with a ferromagnetic coupling $J>0$, which we set to unity throughout, and where $\hat\sigma^i_\rr,\,i\in\{x,y,z\}$, are the Pauli operators at site $\rr$. This model, on square lattices and other geometries, is a paradigm of quantum phase transitions in \cite{sachdev2011quantum,Hamer2006CriticalBehavior,Ishizuka2011QMC,Baek2011QMC,Schmitt2022QuantumPhaseTransition} and out \cite{hashizume2022,Osborne2025Probing} of equilibrium, and is emblematic in investigations of quantum many-body dynamics \cite{Halimeh2025QuantumSimulationOutofequilibrium}.
In the following, we measure all energy (time) scales relative to $J$ ($J^{-1}$). For $\hpa=0$, the Hamiltonian has a global $\mathbb{Z}_2$ spin-flip symmetry which in the ordered phase ($\hpe$ below the quantum critical point $\hpe^\text{c} \approx 3.04$ \cite{duCrooDeJongh1998CriticalBehavior,Bloete2002ClusterMonteCarlo}) is spontaneously broken, resulting in a doubly degenerate ground state. This symmetry is explicitly broken by a finite longitudinal field, lifting the degeneracy. In this case, it is possible to interpret the unique ground state as the ``true'' vacuum of the system, and its counterpart as a ``false'' vacuum. If a system is initialized in the false vacuum, which we define here to be the $\ket{\uparrow}^{\otimes N}$ product state, it is understood that over extremely long timescales vacuum decay can occur through a mechanism involving high-order resonating processes which result in the formation of bubbles of the true vacuum, consisting of down-spins, with a domain wall separating the two configurations.

In continuum field theories, this type of setup is usually studied in a long-wavelength limit via the theory of coarsening~\cite{bray_1994}. In this continuum scenario and for non-conserved order parameter and local interactions, switching on a bias field favoring the true vacuum drives the growth or shrinking of bubbles of the favored configuration. The ultimate fate of an initial bubble is determined by its mean curvature (reciprocal of the radius for a circular bubble) at nucleation, and no other details.\footnote{This assumes that the bubble is ``not too non-convex,'' so that the curvature is always much larger than microscopic scales such as the interface width. It also assumes that the bubble is small compared to system size for a closed system.} In practice, this means that the if the linear length scale of a bubble is $L$, then there is a critical $L_\text{c}\sim 1/\hpa$~\cite{pizzi_apparent_2025} such that bubbles with $L<L_\text{c}$ will shrink and disappear, while those with $L>L_\text{c}$ will grow. This $L_\text{c}$ is determined by the competition between surface tension and bulk energy gain due to the bias field $\hpa$. However, this type of analysis misses lattice effects which, as we will see, cannot be ignored: initial bubbles of very similar shapes rotated with respect to each other behave differently, proving that the shape and orientation of the shape matters, in a way that remains important for all droplet sizes.

\begin{figure}[t!]
    \centering
    \includegraphics[width=\linewidth]{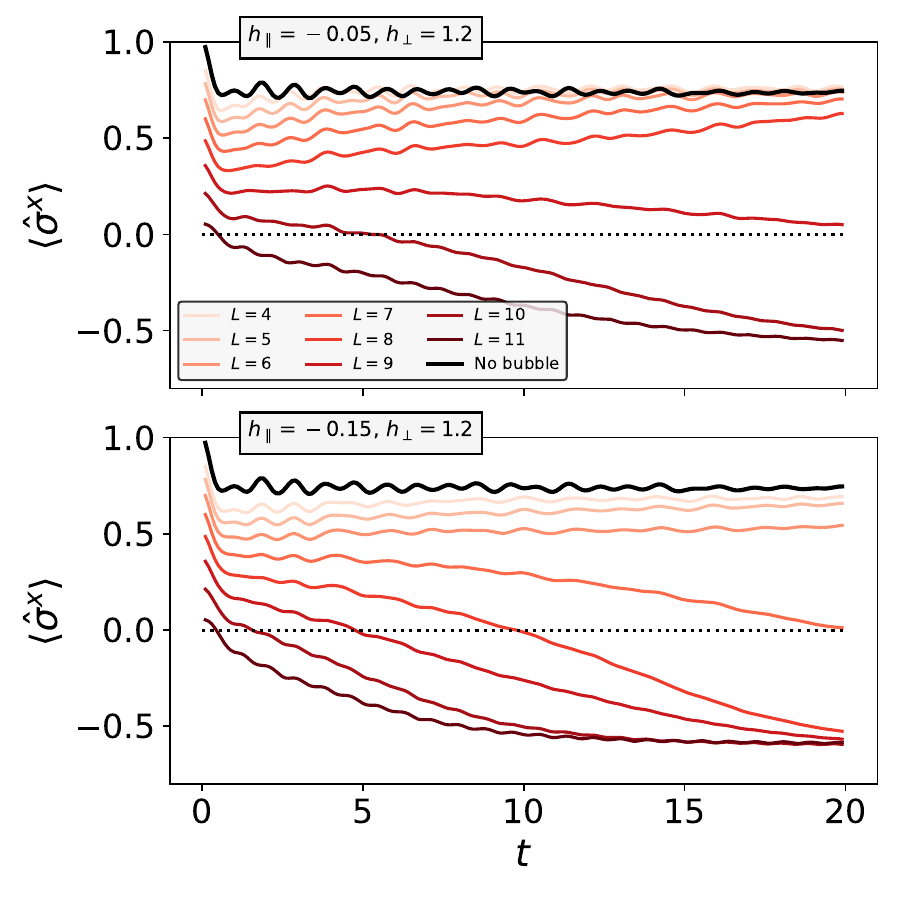}
    \caption{Average value of the magnetization over time after a quench where initial states consisting of square spin-down domains of side $L$ are evolved under the Hamiltonian~\eqref{eq:H_TLFI}. The plots clearly reveal a critical size of the bubble, which, for a fixed value of $\hpe$, depends on $\hpa$. Above this critical size, the bubble expands until it occupies the whole system.}
    \label{fig:crit_L}
\end{figure}

\textbf{\textit{Quench protocol.---}} We focus on a scenario where bubbles of different shapes and sizes are present in the initial state of the system at $t=0$, and we investigate under which conditions they expand or shrink. To numerically simulate the evolution of true-vacuum bubbles in a false-vacuum background, we employ a TTN representation of the wave function \cite{tagliacozzo_simulation_2009, murg_simulating_2010, SM} and apply the time-dependent variational principle (TDVP) \cite{Haegeman2011TDVP,Haegeman2016UnifyingTimeEvolution,Vanderstraeten2019Tangent} using the library $\tt{qtealeaves}$ \cite{qtealeaves}. We start from an initial product state consisting of a spin-down domain (true-vacuum bubble) in a spin-up false-vacuum background, on a $16\times16$ lattice, and quench it with Hamiltonian \eqref{eq:H_TLFI}.

\begin{figure}[t!]
    \centering
    \includegraphics[width=\linewidth]{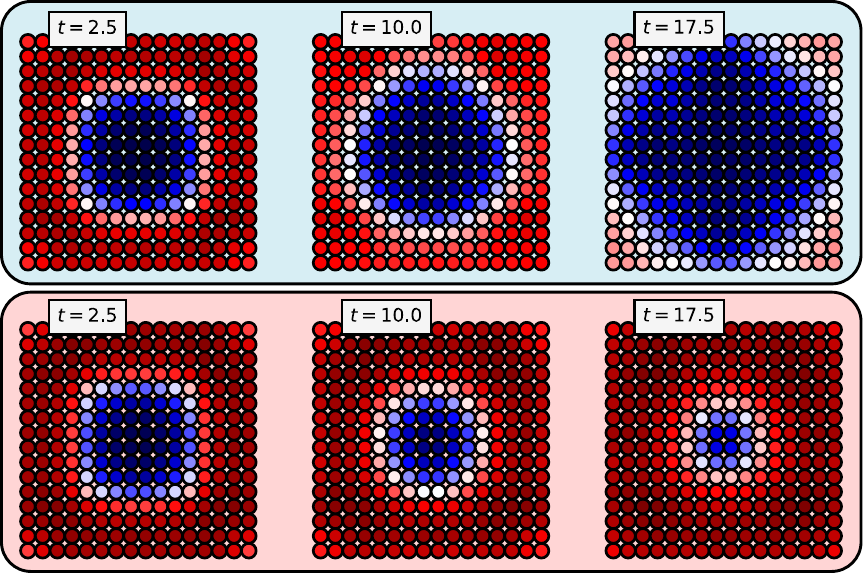}
    \caption{Snapshots of the local magnetization on the $16 \times 16$ lattice at different times for $\hpa=-0.15$ (upper panel) and $\hpa=-0.05$ (lower panel), for an initial $8 \times 8$ square bubble. The change in the longitudinal field drastically affects the bubble dynamics, leading to vacuum decay in the first case and shrinking in the latter. When the bubble expands the time evolution produces diamond-like shapes, which are the square-lattice equivalent of circles.}
    \label{fig:lattices}
\end{figure}

As a first step, we investigate the critical linear size needed for a square-shaped true-vacuum bubble to become critical. To this end we fix the value of the transverse field to $\hpe=1.2$, which is well below the quantum critical point $\hpe^\text{c}$ as well as the dynamical critical point $\hpe^\text{d}\approx2.0$ \cite{hashizume2022}, at which the long-time steady state has zero magnetization for $\hpa=0$. This value of $\hpe=1.2$ is large enough to induce non-perturbative effects over accessible timescales, but small enough so that the background oscillations that are introduced do not hinder the physics that we want to probe \cite{SM}. The fate of this initial state can be tracked by looking at the total average magnetization of the system, $\langle\hat{\sigma}^x\rangle=\sum_\rr\langle\hat{\sigma}^x_\rr\rangle/N$ with $N=256$ the total number of spins, which evolves towards large positive values if the bubble shrinks, and towards large negative values if the bubble expands until it occupies the whole lattice. In the case of square bubbles, the results shown in \cref{fig:crit_L} suggest that $L_\text{c}=7$ at $\hpa=-0.15$ and $L_\text{c}=9$ at $\hpa=-0.05$. In both cases, snapshots of the local magnetization shown in \cref{fig:lattices} for $L=8$ at both these values of $\hpa$ reveal intriguing interface dynamics, where the contraction ($\hpa=-0.05$) or expansion ($\hpa=-0.15$) is accompanied by relaxation towards diamond-like shapes. 
At long times, and for $\hpa \ll J,\hpe$, we expect the absolute value of the average magnetization to approach the value that it would take in the absence of an initial bubble, with its sign depending on whether the true vacuum has taken over (negative) or not (positive). Videos of this dynamics are also provided \cite{videos}.

\begin{figure}[t!]
    \centering
    \includegraphics[width=\linewidth]{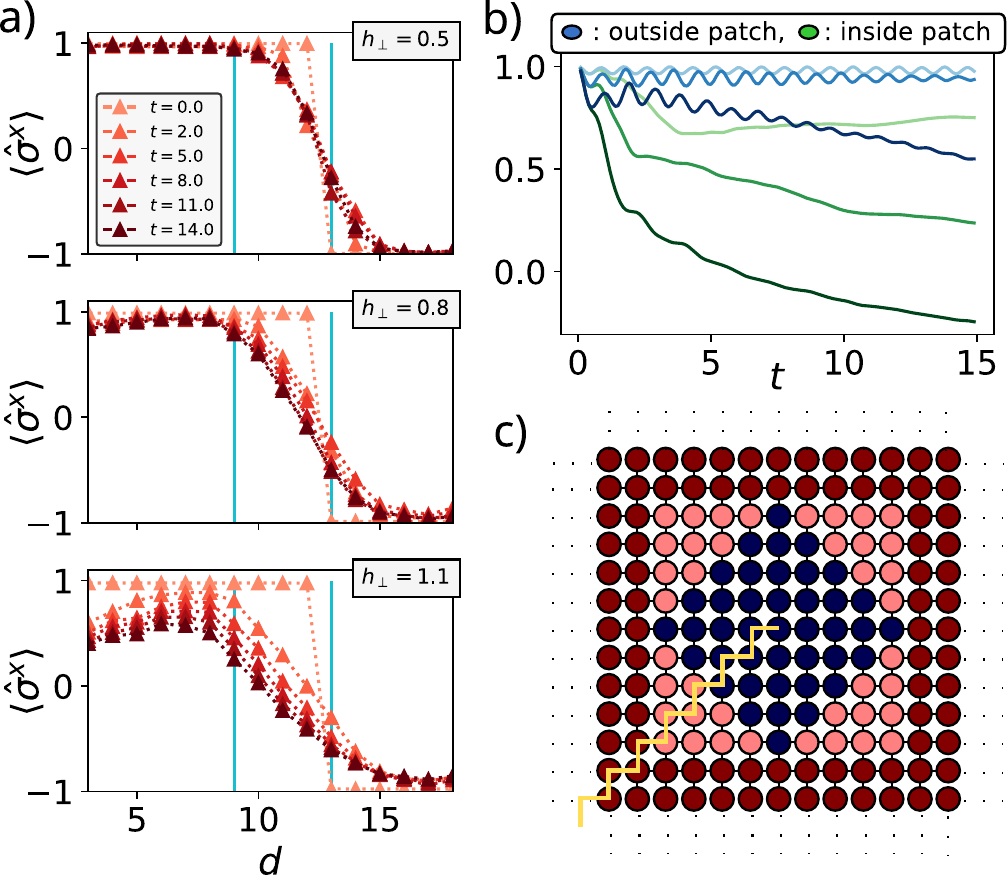}
    \caption{(a) Snapshots of the magnetization along the path shown in panel (c) for different times, $\hpa=-0.25$, and increasing values of the transverse field $\hpe$. (b) Local magnetization at sites inside the square patch (green) and outside of it (blue) for increasing values of $\hpe=0.5, 0.8, 1.1$ (lighter to darker shades). The results show that for moderate values of the transverse field, over intermediate timescales, the dynamics occurs entirely within the patch, in agreement with the perturbative analysis outlined in the main text. (c) Sketch of the system, showing the initial ``true vacuum'' bubble (blue) and the patch covering a domain with the same perimeter as the bubble and maximal area (light red).}
    \label{fig:perp}
\end{figure}

\textbf{\textit{Shape dependence.---}}
We now investigate how the long-time dynamics depends on the shape of the initial bubble. We observe that for a given area, diamonds appear to be a lot more susceptible to expansion. In the limit $\hpe\ll J$, this can be understood in terms of the effective dynamics which preserves the total length of the domain  (perimeter). Indeed, in this regime the initial domains can only increase their area through corner-flipping processes that do not alter the length of the boundary. 
This implies for instance that bubbles featuring many corners, such as diamonds, can evolve into shapes of larger area, the largest being the square in which they can be encased. In this limit, therefore, there is an approximate equivalence between all shapes with the same perimeter, regardless of their area, in that their motion is restricted to the minimal rectangular patch that includes them. The introduction of a longitudinal field, in this context, favors the shapes with a larger area among the ones which are allowed. For the non-perturbative regime that we consider, the above analysis does not apply exactly, but we still expect this mechanism to play an important role. To probe the above argument, in \cref{fig:perp} we quench a diamond-shaped initial bubble for increasing values of the transverse field $\hpe$ at fixed value of the longitudinal field $\hpa=-0.25$, and track the local magnetization. Both spatially resolved snapshots and the time dependence of the magnetization at specific sites, inside and outside of the square patch surrounding the diamond, reveal that up to $\hpe \approx 0.8$ the dynamics is restricted to the patch, in agreement with the above analysis. Above this value, the dynamics extends outside the patch.

\begin{figure}[t!]
    \centering
    \includegraphics[width=\linewidth]{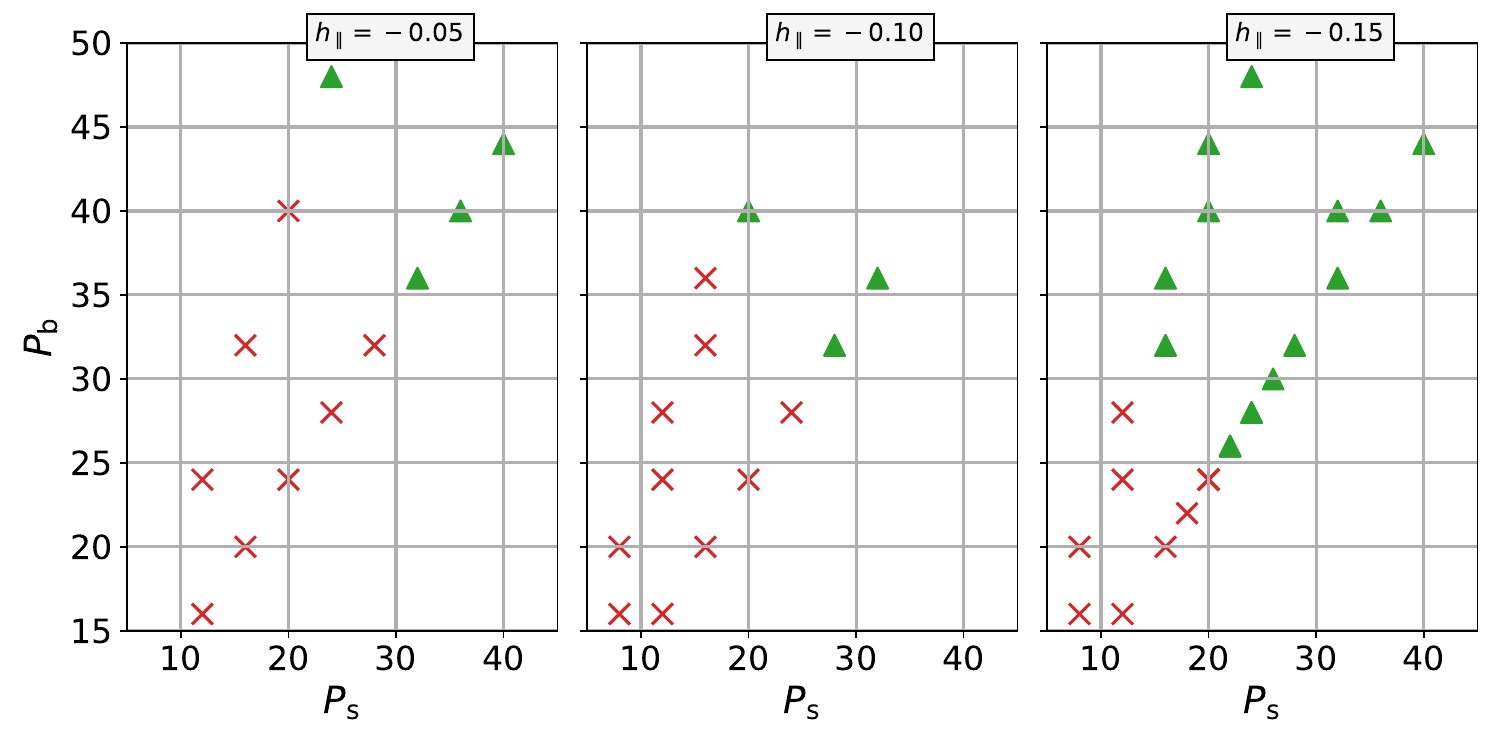}
    \caption{Data points corresponding to initial true vacuum domains of different shape and size, classified exclusively by their bond and site perimeters $P_\text{b}$ and $P_\text{s}$, for longitudinal fields $\hpa={-}0.05,\,{-}0.1,\,\text{and}\,{-}0.15$. Green markers correspond to expanding bubbles. }
    \label{fig:scatter}
\end{figure}

We now address the question of which geometric properties (area, perimeter, curvature, and so on) affect the fate of initial droplets. To this end, we introduce two distinct notions for the perimeter of a droplet. The ``bond perimeter'' $P_\text{b}$ counts the number of broken bonds at the interface, and is therefore related to the energy cost of the domain wall. The ``site perimeter'' $P_\text{s}$, on the other hand, is given by the number of outermost sites of the bubble, so that for instance a square and the corresponding $45^\circ$-rotated diamond have the same $P_\text{s}$ but different $P_\text{b}$. In general, the difference between the two quantities is related to lattice features, as $P_b$ is maximized by shapes with a large number of corners. In \cref{fig:scatter}, we collect the data for a number of shapes (squares, rectangles, diamonds, crosses, and some more irregular ones \cite{SM}) and assign each to a point in the $(P_\text{s},P_\text{b})$ plane, classifying them depending on whether they expand (green) or not (red) after the global quench described above. From the scatter plots, we see that neither $P_\text{b}$ or $P_\text{s}$ on their own are sufficient to characterize the behavior (had this been the case, a vertical or horizontal line would have separated the two sets). Indeed, rotated versions of the same bubble (such as a diamond and a square) may well behave differently. This is in stark contrast to the predictions of Landau-Ginzburg theory, which says that the linear dimension of the droplet is what determines its fate.

\textbf{\textit{Experimental realization.---}} 
The phenomenology discussed in this work can be probed on current quantum simulation platforms, in particular with Rydberg atom arrays.
This platform allows for the control of large enough $2$d arrays of atoms, the implementation of an Ising system, and the required local control.
The Ising Hamiltonian arises if a non-interacting atomic ground state is laser-coupled to a Rydberg state, where the choice of the state determines the interactions.
The laser parameters (Rabi frequency and detuning) set the transverse and longitudinal fields, respectively.
The Rydberg atom interactions decay with distance as $V(r) \propto r^{-6}$, leading to dominating nearest-neighbor couplings with the diagonal interactions being $1/8$ weaker than the nearest-neighbor ones.
We expect that the presence of such longer-ranged interactions will not qualitatively alter the bubble dynamics discussed above.
We foresee two possible routes to implement the setting described above depending on the Rydberg interactions being ferro- or antiferromagnetic.
Both require local laser addressing to prepare the initial state from the natural starting point, the fully polarized (all-up) state.
In the ferromagnetic case, only the bubble region needs to be addressed and spin flipped.
This can be achieved by inducing a dominating light shift of the ground-Rydberg transition and then applying a $\pi$-pulse breaking the Rydberg blockade.
While strong spin-position coupling complicates this preparation, recent experiments have demonstrated the ability to overcome this challenge \cite{osterholz_collective_2025}.
The subsequent dynamics require to change the laser parameters quickly, typically on the few-nanosecond timescale, which can be achieved with current technology.
The bubble dynamics happen on intermediate timescales of a few tens of $J^{-1}$. 
Probing the Rydberg many-body system at these times is challenging and may require trapping of the Rydberg atoms, but it is realistically achievable in the near future.
In the antiferromagnetic case, the energy of the two degenerate ground states can be altered by introducing a staggered longitudinal field by locally shifting the energy of the Rydberg levels~\cite{darbha_false_2024, chao_probing_2025}.
The opposite staggered field is then applied in the bubble region.
The subsequent $\pi$-pulse to create the bubble and background state also needs to break the Rydberg blockade at the interface between the two regions.
The dynamics can then be initiated by changing the laser parameters as in the ferromagnetic case.

\textbf{\textit{Summary and outlook.---}} 
We have presented a numerical TTN study of bubble dynamics in the $2+1$D quantum Ising model. Our results clearly show how the geometric features of the initial domain and the microscopic parameters of the Hamiltonian determine drastically different dynamical regimes, corresponding to expansion (vacuum decay) or contraction of the bubble, in a way that cannot be inferred from simple continuum theories. The evolution of the average magnetization of the system provides a clear signature of such behavior over intermediate timescales, which are within reach for Rydberg array-based quantum simulators. To this end, we have proposed a realistic quantum simulation scheme aimed at observing false-vacuum decay related phenomenology on currently available quantum hardware. 

The ability to study real-time quench dynamics for relatively large $2+1$D systems in non-perturbative regimes and far from equilibrium, as showcased in our Letter, paves the way for several future extensions of our work. A particularly relevant direction is the generalization of our protocol to lattice gauge theories, where false vacuum decay is particularly relevant from a high-energy physics point of view \cite{Gattringer2009QuantumChromodynamicsLattice,Weinberg1995QuantumTheoryFields,Zee2003QuantumFieldTheory,Ellis2003QCDColliderPhysics}, and where quantum simulation experiments of these models are being performed on various types of quantum hardware \cite{Martinez2016RealtimeDynamicsLattice, Klco2018QuantumclassicalComputationSchwinger,Gorg2019RealizationDensitydependentPeierls, Schweizer2019FloquetApproachZ2, Mil2020ScalableRealizationLocal, Yang2020ObservationGaugeInvariance, Wang2022ObservationEmergent$mathbbZ_2$, Su2023ObservationManybodyScarring, Zhou2022ThermalizationDynamicsGauge, Wang2023InterrelatedThermalizationQuantum, Zhang2025ObservationMicroscopicConfinement, Zhu2024ProbingFalseVacuum, Ciavarella2021TrailheadQuantumSimulation, Ciavarella2022PreparationSU3Lattice, Ciavarella2023QuantumSimulationLattice-1, Ciavarella2024QuantumSimulationSU3, 
Gustafson2024PrimitiveQuantumGates, Gustafson2024PrimitiveQuantumGates-1, Lamm2024BlockEncodingsDiscrete, Farrell2023PreparationsQuantumSimulations-1, Farrell2023PreparationsQuantumSimulations, 
Farrell2024ScalableCircuitsPreparing,
Farrell2024QuantumSimulationsHadron, Li2024SequencyHierarchyTruncation, Zemlevskiy2025ScalableQuantumSimulations, Lewis2019QubitModelU1, Atas2021SU2HadronsQuantum, ARahman2022SelfmitigatingTrotterCircuits, Atas2023SimulatingOnedimensionalQuantum, Mendicelli2023RealTimeEvolution, Kavaki2024SquarePlaquettesTriamond, Than2024PhaseDiagramQuantum, Angelides2025FirstorderPhaseTransition, Gyawali2025ObservationDisorderfreeLocalization, Cochran2025VisualizingDynamicsCharges, Gonzalez-Cuadra2025ObservationStringBreaking, Crippa2024AnalysisConfinementString, De2024ObservationStringbreakingDynamics, Liu2024StringBreakingMechanism, Alexandrou2025RealizingStringBreaking, 
Mildenberger2025Confinement$$mathbbZ_2$$Lattice, Schuhmacher2025ObservationHadronScattering, Davoudi2025QuantumComputationHadron, Cobos2025RealTimeDynamics2+1D, Saner2025RealTimeObservationAharonovBohm, Xiang2025RealtimeScatteringFreezeout, Wang2025ObservationInelasticMeson,li2025frameworkquantumsimulationsenergyloss,froland2025simulatingfullygaugefixedsu2,Hudomal2025ErgodicityBreakingMeetsCriticality}. 

\bigskip
\footnotesize
\begin{acknowledgments}
\textbf{\textit{Acknowledgments.---}}U.B.~and J.C.H.~acknowledge funding by the Max Planck Society, the Deutsche Forschungsgemeinschaft (DFG, German Research Foundation) under Germany’s Excellence Strategy – EXC-2111 – 390814868, and the European Research Council (ERC) under the European Union’s Horizon Europe research and innovation program (Grant Agreement No.~101165667)—ERC Starting Grant QuSiGauge. Views and opinions expressed are, however, those of the author(s) only and do not necessarily reflect those of the European Union or the European Research Council Executive Agency. Neither the European Union nor the granting authority can be held responsible for them. This work is part of the Quantum Computing for High-Energy Physics (QC4HEP) working group. A.L.~acknowledges support from the Leverhulme Trust Research Project Grant RPG-2025-063. C.G.~acknowledges support from the Horizon Europe program HORIZON-CL4-2022-QUANTUM-02-SGA via the project 101113690 (PASQuanS2.1), the Deutsche Forschungsgemeinschaft through FOR 5522 (Grant No.~499180199), and the Alfried Krupp von Bohlen and Halbach Foundation.
\end{acknowledgments}
\normalsize

\bibliography{biblio,biblio_LGT}

\clearpage
\pagebreak
\newpage

\onecolumngrid

\setcounter{equation}{0}
\setcounter{figure}{0}
\setcounter{table}{0}
\setcounter{page}{1}

\renewcommand{\theequation}{S\arabic{equation}}
\renewcommand{\thefigure}{S\arabic{figure}}
\renewcommand{\thetable}{S\arabic{table}}

\begin{center}
\textbf{\large Supplemental Material}
\end{center}
\section{Numerical methods and convergence}
To numerically investigate the quench dynamics, we rely on a Tree Tensor Network (TTN) representation of the wave function, which is evolved in time using the Time Dependent Variational Principle (TDVP). TTNs are a generalization of Matrix Product States (MPS) whose building block are tensors of rank $3$ arranged into a binary tree, whose lower layer contains the physical legs. Due to their higher connectivity, in TTNs the distance between two physical sites within the network is always logarithmic with respect to their distance on the 1d lattice. This is particularly relevant when adapting methods which are inherently one-dimensional to two-dimensional lattices, as this inevitably comes at the cost of introducing artificially long-range interactions. For the simulations we use the $\texttt{qtealeaves}$ library \cite{qtealeaves}, which implements TTNs, an efficient mapping of two-dimensional lattices to one-dimensional Hilbert curves, and the a variety of tensor-network based algorithms. Numerical simulations are performed using the single-site TDVP algorithm, and are significantly sped up by the use of GPUs. For all the figures shown, we ran simulations on either NVIDIA H100 NVL GPUs with 92GB of RAM, or on NVIDIA A100 GPUs with 80GB of RAM, with bond dimension up to $\chi=256$ and timestep $dt=0.05$.  

\subsection{Convergence}
\begin{figure}[htp]
    \centering
    \includegraphics[width=0.8\linewidth]{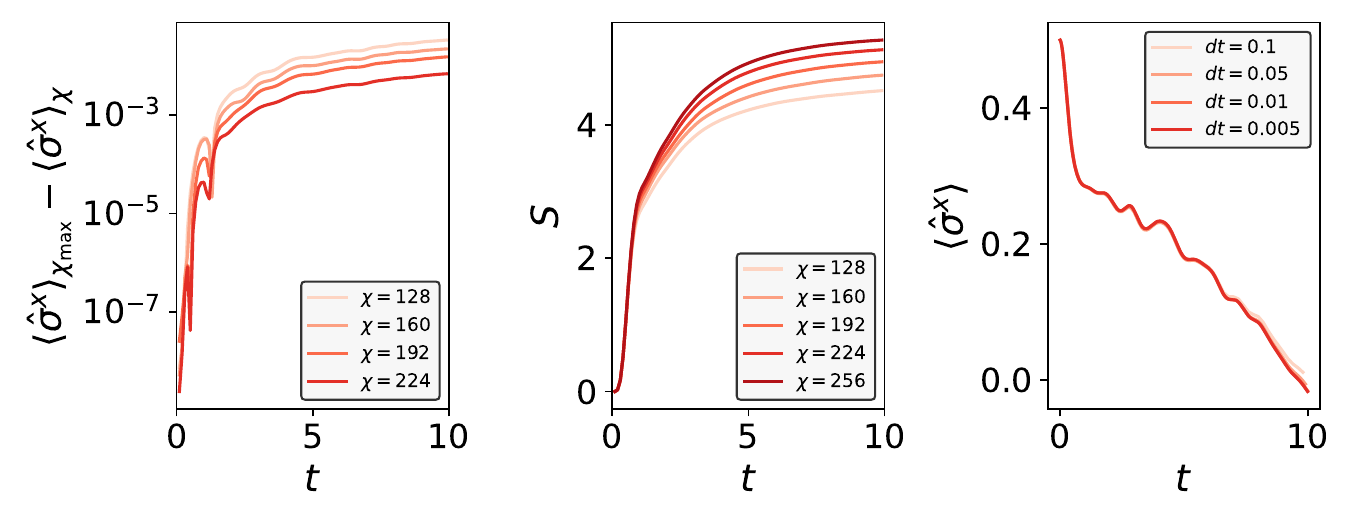}
    \caption{Convergence tests for a quench with parameters $\hpe=1.2$ and $\hpa=-0.15$ and an initial state corresponding to a square bubble of side $L=8$.}
    \label{fig:convergence}
\end{figure}

As a test for the convergence of the single-site TDVP algorithm, we perform the same quench of an $8\times 8$ square bubble for different values of the bond dimension $\chi$ and the time step $dt$. The transverse and longitudinal fields are fixed to $\hpa=-0.15$ and $\hpe=1.2$. As shown in \cref{fig:convergence}, the change in local magnetization is of order $10^{-3}$ between the two largest values of $\chi$, $256$ and $224$. As expected, a larger bond dimension allows to capture more entanglement. Due to the finite size of the system, in this quench the bubble expansion reaches the boundary at relatively short times, after which only oscillations in the magnetization are visible. 

We also point out that, since the TTNs are inherently one-dimensional, they do not encode in any way the $C_4$ symmetry of the square lattice. When starting from a symmetric configuration (e.g a $6\times 6$ square at the center of a $16 \times 16$ lattice) we expect this symmetry to be preserved by the time evolution. The fact that this is the case, as visible for example in \cref{fig:lattices}, provides a further test of the convergence of the algorithm.

\section{Dependence on the transverse field}
\begin{figure}[t!]
    \centering
    \includegraphics[width=0.5\linewidth]{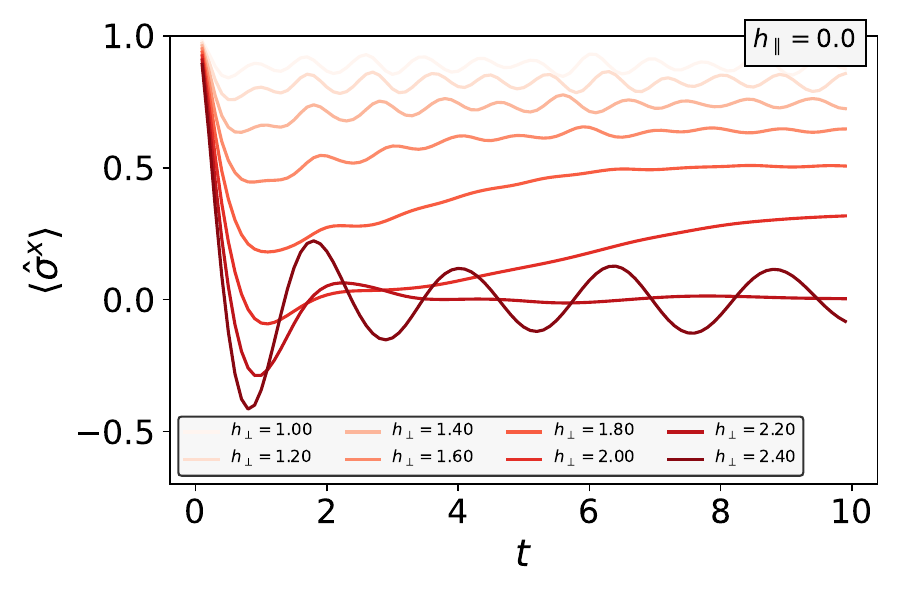}
    \caption{Dependence of the magnetization on the transverse field $\hpe$ for an homogeneous quench of the pure ferromagnetic state $\ket{\uparrow}^{\otimes N}$ at zero longitudinal field. A finite transverse field causes oscillations of the background and an overall reduction in the average magnetization over time. We observe that for $\hpe \gtrapprox 2.0$ the magnetization goes to zero at long times.}
    \label{fig:h_transv_dep}
\end{figure}

In the main text, we have fixed the transverse field to $\hpe=1.2$ everywhere except for \cref{fig:h_transv_dep}. This value corresponds to a non-perturbative regime $\hpa \approx J$ where little analytical insight can be obtained. At the same time, a too large transverse field would cause strong oscillations of the background, making it difficult to detect the dynamics of the bubbles. In \cref{fig:h_transv_dep} we show how the average background magnetization oscillates in the absence of a longitudinal field. The results show that the amplitude, frequency and overall trend of the oscillations depend greatly and not always monotonically on $\hpe$. In particular, we find that for values of the transverse field $\hpe \gtrapprox 2.0$ the average magnetization of the system drops to zero at long times. This fully agrees with the detection of a dynamical quantum critical point at this value of the transverse field in \cite{hashizume2022}.

\section{Initial states}
\begin{figure}[t!]
    \centering
    \includegraphics[width=0.8\linewidth]{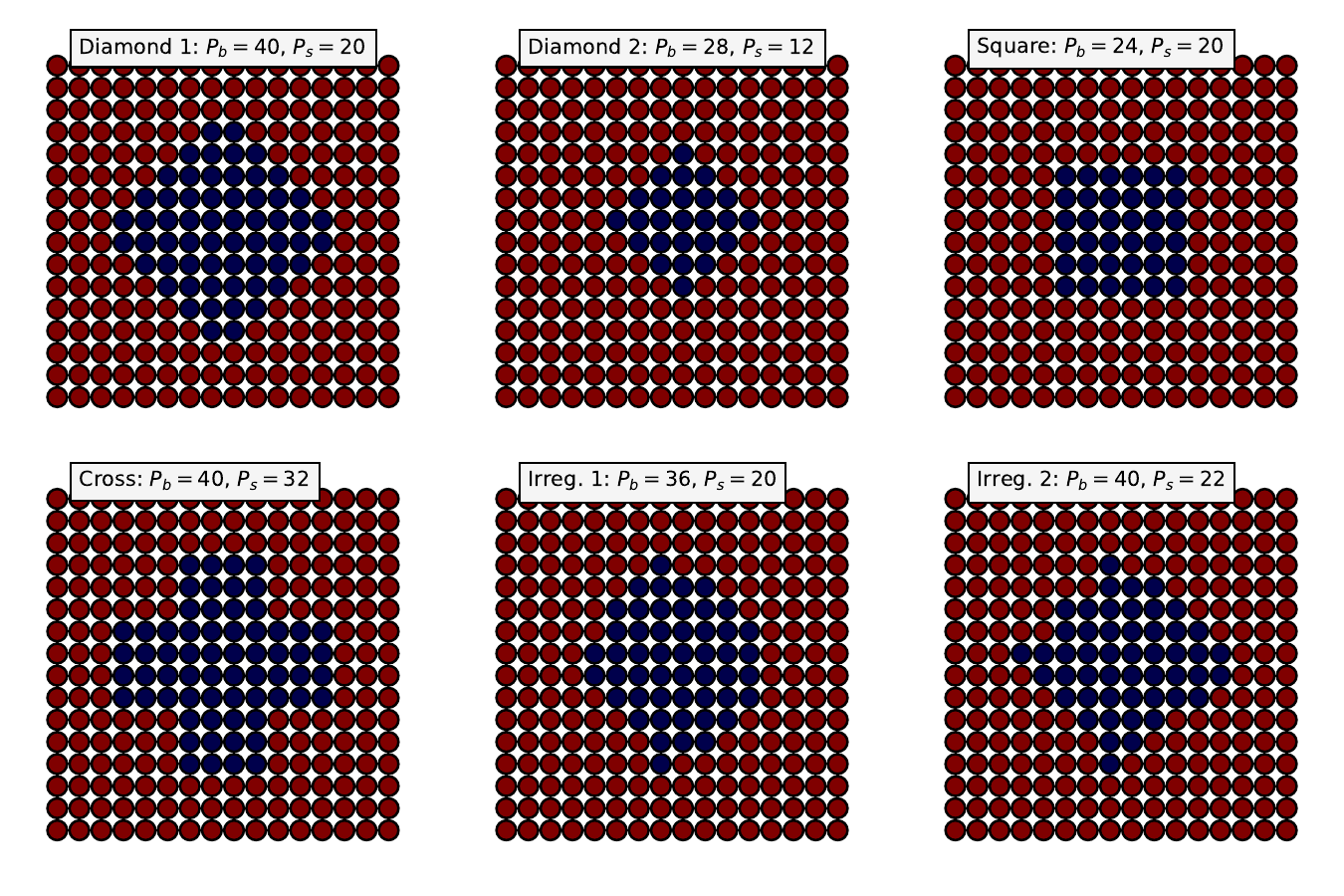}
    \caption{Examples of bubble shapes that are used as initial states for the homogeneous quenches discussed in the main text. Together with the configuration, in each panel we indicate the value of the ``bond'' and ``site'' perimeters $P_\text{b}$ and $P_\text{s}$ defined in the main text.}
    \label{fig:init_states}
\end{figure}

For the quench protocol outlined in the main text we initiate the system in a number of different initial states, as we are interested in how the bubble dynamics depends on the geometry of the domain. We show examples of such initial states in \cref{fig:init_states} and, in each case, we indicate the value of the ``bond'' and ``site'' perimeters $P_\text{b}$ and $P_\text{s}$ as defined in the main text.

\end{document}